# Spin State of Single Molecular Magnet (SMM) Creating Long Range Ordering on Ferromagnetic Layers of Magnetic Tunnel Junction -A Monte Carlo Study

Andrew Grizzle, Christopher D'Angelo, Pawan Tyagi*
Center for Nanotechnology Research and Education, Mechanical Engineering, University of the District of Columbia, Washington DC-20008, USA
Email of corresponding author: ptyagi@udc.edu

**ABSTRACT** Single molecular magnet (SMM) like paramagnetic molecules interacting with the ferromagnetic electrodes of a magnetic tunnel junction (MTJ) produce a new system that differs dramatically from the properties of isolated molecules and ferromagnets. However, it is unknown how far deep in the ferromagnetic electrode the impact of the paramagnetic molecule and ferromagnet interactions can travel for various levels of molecular spin states. Our prior experimental studies showed two types of paramagnetic SMMs, the hexanuclear Mn6 and octanuclear Fe-Ni molecular complexes, covalently bonded to ferromagnets produced unprecedented strong antiferromagnetic coupling between two ferromagnets at room temperature leading to a number of intriguing observations. In this paper, we report Monte Carlo Simulations (MCS) study focusing on the impact of the molecular spin state on cross junction shaped MTJ based molecular spintronics device (MTJMSD). Our MCS study focused on the Heisenberg model of MTJMSD and investigated the impact of various molecular coupling strengths, thermal energy, and molecular spin states. To gauge the impact of the molecular spin state on the region of ferromagnetic electrodes, we examined the spatial distribution of molecule-ferromagnet correlated phases. Our MCS study shows that under a strong coupling regime molecular spin state should be ~30% of the ferromagnetic electrode's atomic spins to create long-range correlated phases.

## I. INTRODUCTION

Molecules are the only mass-producible nanostructures with customizable chemical, electrical, optical, and magnetic properties that can be prod with sub-angstrom scale precision. Molecules are extremely versatile and practically billions of types are possible and so are the molecule-based devices[1-3]. Several molecules such as single molecular magnets (SMM)[4], porphyrin[5], DNA[6], organometallic molecules[7] have a high potential to be included as the device element in futuristic molecular spintronics devices (MSD). MSD fabrication requires a molecule of interest to be simultaneously connected with at least a source and drain-type metal electrode[8]. The intensity of interaction can be weak if it is physically separated from the two-metal electrode or connected by weak bonds[9]. However, a molecule with functional groups like sulfur can form covalent and ionic bonds leading to very strong coupling[10, 11]. In the strong coupling regime, molecules and metal electrodes near the interface show strong hybridization of energy levels[12]. Such strong hybridization has been observed to create novel properties on both metal electrodes and molecules. For example, the interaction of thiolate molecule produced magnetism in non-magnetic electrode[13] and further enhanced the degree of spin polarization on ferromagnets. It is also well known that a molecule connected to metal electrodes cannot exhibit the properties measured in its isolated state. Therefore, the combined system of metal electrodes and molecules becomes a new composite system altogether[13, 14]. Understanding this system is extremely important to progress the field of MSD where SMM-like molecules possess a wide range of spin states interacting with magnetic electrodes[13]. Magnetic electrodes, such as nickel (Ni), cobalt (Co), iron (Fe), exhibit strong long-range ordering. This long-range ordering can further transport the effect of molecule-ferromagnet interaction over microscopic range. Our previous experimental studies showed that Mn hexanuclear[15] and Fe-Ni octanuclear molecular complexes(OMC)[14] based SMM produced long-range impacts on ferromagnetic electrodes leading to room temperature observations of several orders current suppression, spin photovoltaic effects, and several orders of magnitude magnetoresistance[15, 16]. Other groups have also observed strong coupling between $C_{60}$ molecules and ferromagnetism of the nickel electrodes leading to the Kondo splitting phenomenon without applying the estimated ~50 T field needed for this observation[17]. However, experimentally determining the spin state of a paramagnetic molecule after forming a complete MSD is extremely challenging. Additionally, Density Function Theory (DFT) study is extremely challenging to simulate SMM-connected to realistic large-scale MSDs with long ferromagnetic electrodes[18]. This paper investigates the effect of molecular spin state on the experimentally studied cross junction shaped MTJMSDs[15, 16] with extended ferromagnetic electrodes beyond the molecular junction area. For this study, we have employed the Heisenberg Model[19] of MTJMSD that showed promising results in our prior MCS[20]. This paper provides new insights into the effect of molecular spin state and evaluates the properties of whole MSD. In addition, we have varied the molecular spin state from low to high to observe its impact on long-range ordering on the MTJMSD.

## II. METHOD

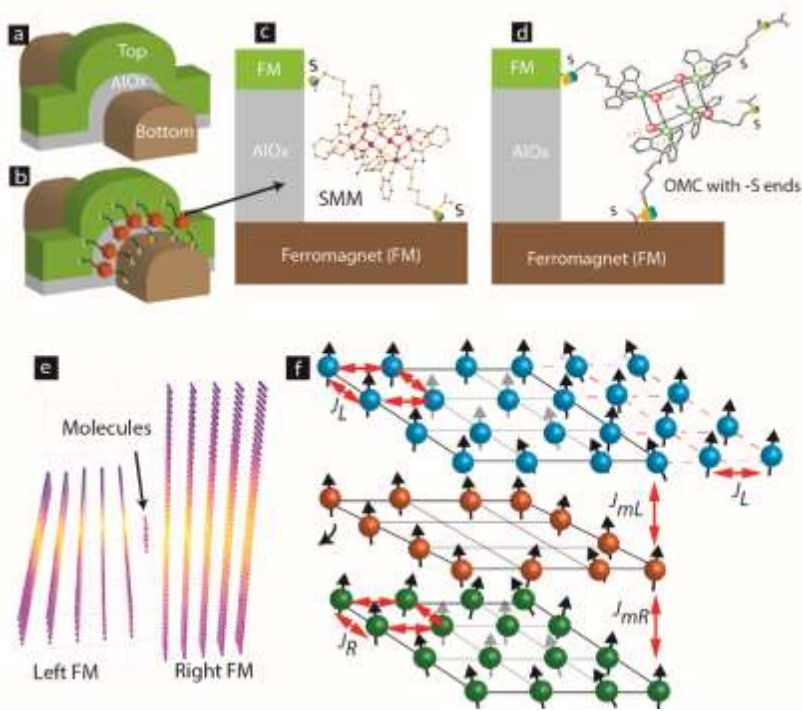

We utilized a continuous spin model to allow spin vectors of the ferromagnets' atoms and molecules to assume any directions in a spherical coordinate system[21]. To understand the property of experimentally studied MTJMSD via this MCS study, we focused on the Heisenberg model (Fig.1e) as a 3D analog of an MTJMSD (Fig. 1b)[14]. This MCS study represented a tunnel barrier with empty space within a square-shaped molecular perimeter (Fig.1f). The molecular perimeter was a 5x5 square with 16 molecular analogs; in Fig.1f, we showed a 4x4 molecular square to produce a vivid illustration. Paramagnetic SMM molecules of MTJMSD (Fig.1d) were represented by the atomic scale analog with adjustable spin ($S_m$) parameter. The coupling between two FM electrodes were only caused by the paramagnetic molecules (Fig. 1f). The molecule-mediated exchange coupling between the left and right FM electrodes is governed by the molecule's coupling with left electrode ($J_{mL}$) and molecule coupling with the right electrode ($J_{mR}$), respectively. To simulate the effect of change in ambient temperature we varied thermal energy ($kT$) of the MTJMSD Heisenberg model. The MTJMSD energy was calculated using equation (1). A new state was selected or rejected according to the Metropolis algorithm[21].

**FIG. 1**. MSD formed by utilizing exposed edges of (a) Bare MTJ to attach (b) paramagnetic molecules between two ferromagnets. (c) SMM and (d) OMC paramagnetic molecules connected to ferromagnets via sulfur atom. (e) 3D Heisenberg Model of Molecular device. (f)) Exchange coupling parameters associated with molecule-ferromagnet interactions.

$$U = -J_L \left(\sum_{i \in L} \vec{S}_i \vec{S}_{i+1}\right) - J_R \left(\sum_{i \in R} \vec{S}_i \vec{S}_{i+1}\right) - J_{mL} \left(\sum_{i \in L, i+1 \in mol} \vec{S}_i \vec{S}_{mi+1}\right) - J_{mR} \left(\sum_{i-1 \in mol, i \in R} \vec{S}_{mi-1} \vec{S}_i\right) (1)$$

In this study, $S$ is a 3D vector that represents the discrete atomic spin of FM electrodes. $S_{mi}$ vectors represent the $S_m$ of molecules at $i^{th}$ position. $S_m$ was varied over the 0 to 4 range. However, main discussion is around the critical $S_m$ values for which transition in the molecular device was observed. $J_L$, and $J_R$, are the Heisenberg exchange coupling strengths for the left and right FM electrodes (Fig. 1b). The molecule's spin was coupled with ferromagnetic electrode spin via the exchange coupling parameters and create a correlated system. In our MCSs, the atoms beyond the boundary of the MTJMSD model (Fig. 1b) were set with zero spin state[21]. Energy ($U$), described in equation (1), of the whole system, was minimized by running the Markov process. Markov process generated a new state after 200-2000 Million iterations to reach a stable low energy state. Further details of MCS are published elsewhere[14]. The units of total energy $U$ and exchange coupling parameters are the same as $kT$. In this study, the exchange coupling parameters and $kT$ are referred to as the unitless parameters. The overall magnetic moment of the MTJMSD is the sum of the magnetic moment of the molecules, Left FM and Right FM electrodes. We have mainly focused on the molecule-induced strong antiferromagnetic coupling where $J_{mL}$= -1 and $J_{mR}$ =1. The reason for the emphasis on the molecule-induced antiferromagnetic coupling is the observation of molecule-induced strong exchange coupling in our prior experimental work[14]. To make this study generic, we also varied molecular coupling strength, thermal energy, molecular spin state, and MTJMSD dimensions.

### III. RESULTS AND DISCUSSIONS

To investigate the impact of $S_m$, we recorded the magnetic moment of the FM electrodes and the MTJMSD as a function of iterations steps. We generally ran a MCS over ~200 Million iterations and recorded the magnetic moment of the FM electrodes, molecules, and whole MTJMSD at the interval of 50,000 steps. According to our previous study, OMC induced strong antiferromagnetic coupling[14]. Since we experimentally observed molecule-induced strong antiferromagnetic coupling well above room temperature[14], we have investigated stabilization study at $kT$=0.1. To represent molecule-induced strong antiferromagnetic exchange coupling, we fixed $J_{mL}$ = -1 and $J_{mR}$ = 1. We varied $S_m$ from 0 to 4 range. However, we observed that the nature of MTJMSD stabilization dramatically changed around $S_m$ = ~0.2 (Fig. 2a). For $S_m \leq 0.1$, the left ferromagnet (Left-FM) and right ferromagnet (Right-FM) stabilized around 1200 magnetic moment (Supplementary Material-Fig.S1). However, MTJMSD stabilized near 2000 (Fig. 2a). For $S_m \geq 0.3$, Left-FM and Right-FM both still stabilized around 1000. However, MTJMSD's total magnetic moment, which is the sum of the magnetic moment of Left-FM, Right-FM, and molecules, started settling below the individual electrode magnetic moment around 600. This result suggests that even though the molecule made the same level of strong coupling with two electrodes but, $S_m$ dictate the MTJMSD stabilization dynamics. We also explored the effect of a wider range of $S_m$ (Fig. 2c) on MTJMSD and left and right FM electrodes. The Left-FM and Right

FM electrodes settled around 1100, i.e., close to their maximum possible magnetic moment of FM electrodes, i.e.,1250 for $S_m$ range from 0 to 1 (Fig. 2c). Interestingly, around $S_m = 0.2$, molecule started forcing Left-FM and Right-FM to settle in the antiparallel state due to the molecule induced antiferromagnetic coupling (Fig. 2c). This result suggests that strong exchange coupling between molecule and FM electrodes can only impact MTJMSD when the molecular spin magnitude is above a critical value. For $S_m = 4$, we saw FM electrode and MTJMSD stabilization pattern like that of $S_m = 1$ (Supplementary Material-Fig.S2). However, major difference was that from a very early stage the MTJMSD magnetic moment became lower than that of Left-FM and Right-FM electrodes. It means increasing $S_m$ promoted early stabilization of MTJMSD into an antiferromagnetic state.

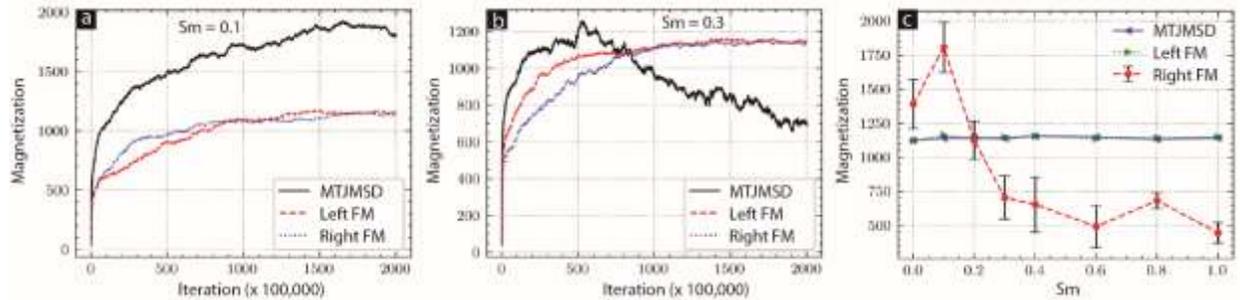

**FIG .2**. Iteration count vs magnetic moment of MTJMSD, left FM, and right FM for (a) $S_m = 0.1$, (b) $S_m = 0.3$, (c) MTJMSD and Fm elecode magnetic moment for molecular spin ranging 0 to 1. For all the cases $kT = 0.1$, $J_{mL} = -1$ and $J_{mR} = 1$.

Understanding the spatial range of $S_m$ on the electrode is critical in understanding how far molecule influence can penetrate FM electrodes. For the calculation of spatial correlation between molecular spin state and the magnetic electrode's spin state, we calculated the dot product between the average magnetic moment of the molecules with each atom's magnetic moment in Left-FM and Right-FM and termed this product as correlation factor. To produce a 2D spatial correlation factor graph, we averaged the data along the width of the FM electrode; here, width is the shorter dimension parallel to the molecular plane (Fig.1e).

We also investigated the spatial magnetic susceptibility of MTJMSD. For the magnetic susceptibility calculation, the magnetic moment of 16 molecules were utilized. However, for the calculation of spatial magnetic susceptibility of the FM electrodes the magnetic moment (m) of each atom present along the width dimensions, shorter dimension parallel to the molecular plane, of each FM electrodes were utilized (Equation 2)[21]

$$\chi = kT.\text{N}(\langle m^2 \rangle - \langle m \rangle^2)$$ --------------(2)

For the case of $S_m = 0.1$, molecules' magnetic susceptibility ($\chi$) was very high as compared to the two FM electrodes (Fig. 4a). A higher $\chi$ for molecule suggests that for $S_m = 0.1$ external magnetic field can align the molecular spin vector selectively. However, for $S_m = 0.3$ case, the magnitude of for molecule were around 4 and 0, respectively (Fig. 4b). For $S_m = 1$, this difference between the $\chi$ for molecules and magnetic electrode were ~1 and 0, respectively (Fig. 4c). Ultimately, for $S_m = 4$, the value of $\chi$ for molecules and FM electrodes was almost the same and near 0 (Fig.4d). This study suggests that if an MTJMSD possesses strongly exchange-coupled high spin molecular magnets, then realizing selective switching of molecules will be extremely challenging.

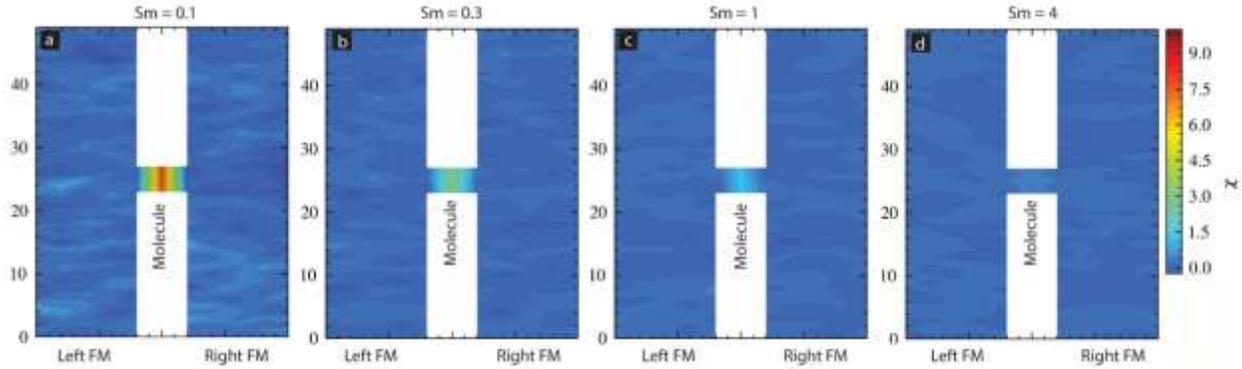

**FIG .4**. Magnetic susceptibility($\chi$) of FM electrodes and molecular layers of MTJMSD for (a) $S_m$ = 0.1, (b) $S_m$ = 0.3, (c) $S_m$ = 1, and (d) $S_m$ = 4. For all the cases $kT$ = 0.1, $J_{mL}$ = -1 and $J_{mR}$ = 1.

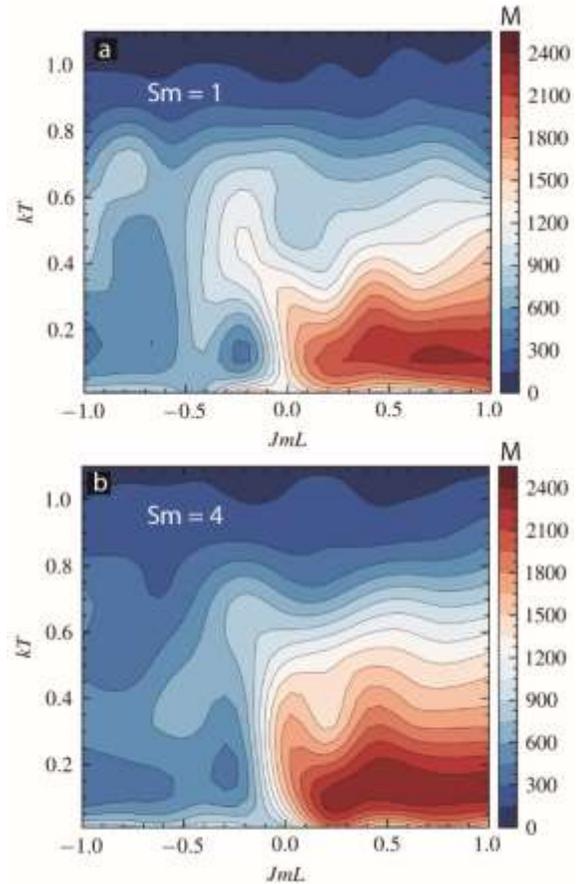

**FIG. 5**. Contour plots of magnetic moment of MTJMSD as a function of $kT$ and $J_{mL}$ & $|J_{mR}|$ for (a) $S_m$ = 1, and (b) $S_m$ = 4.

In the data discussed in figures 2-4 we only discussed $kT$ = 0.1 and $J_{mL}$ = -1 and $J_{mR}$ = 1. To make this study applicable for a wide range of possibilities, we investigated the effect of thermal energy and molecular coupling strengths on MTJMSDs for different $S_m$. To investigate the effect of thermal energy, we varied $kT$ from 0.01 to 1.1. The molecular coupling strength was varied by ensuring that the modulus of $J_{mL}$ and $J_{mR}$ were equal. To consider both cases, molecule inducing ferromagnetic and antiferromagnetic coupling, $J_{mL}$ was varied from -1 to 1 range. In this case $J_{mR}$ was equal to $|J_{mL}|$. The contour plot for $S_m$ = 0 shows that MTJMSD's magnetic moment settled in high and low magnitude state irrespective of the sign and magnitude of $J_{mL}$ and $J_{mR}$ (Supplementary Material-Fig. S4). Increasing $kT$ settled MTJMSD into a highly disordered state producing a low MTJMSD magnetic moment. Contour plot for $S_m$ = 1 and $kT<0.2$ the MTJMSD's magnetic moment remained close to 300-900 for negative $J_{mL}$ and $J_{mR}=|J_{mL}|$ (Fig. 5a). Interestingly, MTJMSD low magnetic moment state was more prevalent on both sides of the $J_{mL}$ = -0.5. As $kT$ increased, the MTJMSD started attaining the higher magnetic moment and finally settled into a low magnetic moment state due to thermal energy induced disordering (Fig. 5a). Contour plot for $S_m$ = 1 and $kT<0.2$ the MTJMSD's magnetic moment was as high as ~2400 for positive $J_{mL}$ and $J_{mR}= J_{mL}$ (Fig. 5a). For a positive sign of $J_{mL}$ and $J_{mR}$, as $kT$ increased, the MTJMSD's magnetic moment started attaining the lower higher magnetic moment and finally settled into a low magnetic moment state due to thermal energy induced disordering

(Fig. 5a). The contour plot for $S_m = 4$ was somewhat similar to that of $S_m = 1$ (Fig. 5b). However, for $S_m = 4$ and $kT < 0.2$, the MTJMSD's magnetic moment persisted around ~500 for weaker molecular coupling. For instance, MTJMSD magnetic moment state that was seen for $S_m = 1$ around $kT = 0.1$-$0.2$ for $J_{mL} \leq -0.9$ was seen for $S_m = 4$ around $kT = 0.1$-$0.2$ for $J_{mL} \leq -0.6$ (Fig. 5b). Also, MTJMSD magnetic moment state that was seen for $S_m = 1$ around $kT = 0.1$-$0.2$ for over very tight space for positive $J_{mL}$ (0.6-1) was seen over a broad range for $S_m = 4$ around $kT = 0.1$-$0.2$ for $0.2 \leq J_{mL} \leq 1$ (Fig. 5b). $S_m$ played an important role in deciding the overall MTJMSD magnetic moment.

We also investigated the effect of $S_m$ and thermal energy on various parts of the MTJMSDs (Fig. 6). For this study, we focused on $S_m$ ranging from 0 to 0.4 and $kT$ ranging from 0.01 to 0.5 for $J_{mL} = -1$ and $J_{mR} = 1$. The ranges of $S_m$ and $kT$ is selected to focus on major transitions observed in Fig. 2 and Fig. 5. In the contour plot of MTJMSD's magnetic moment was ~2000 for $S_m < 0.2$ and $kT < 0.1$ (Fig. 6a). However, as $S_m$ goes beyond 0.2, MTJMSD started to settle in the low magnetic moment state due to molecule-induced strong antiferromagnetic coupling (Fig. 6a). This result is congruent with the data shown in Fig. 2c. It is important to note that with increasing $kT$, for $S_m < 0.2$, MTJMSD loses a high magnetic moment state very rapidly as compared to the variations observed for $S_m > 0.2$ (Fig. 6a). It is apparent that MTJMSD magnetic moment starts to get coupled with molecular spin state for $S_m > 0.2$, which remains stable for higher thermal energy. The molecule's cumulative magnetic moment also gets impacted due to $kT$ (Fig. 6b). The net magnetic moment of the molecule got disturbed with a slight increase in $kT$ (Fig. 6b). However, as $kT$ increases the molecular magnetic moment persisted more for the higher magnitude of $S_m$. However, Left-FM (Fig. 6c) and Right-FM (Fig. 6d) both showed high magnetic moment for $kT < 0.2$ over 0-0.4 molecular spin magnitude. Electrode finally settled into a thermally induced disturbed low magnetic moment state (Fig. 6c-d). The main message this study suggests is that uniform molecular magnetic moment existed around linear boundaries on $S_m$ vs $kT$ graph (Fig. 6b).

We also investigated the effect of MTJMSD's along with $S_m$. For this study, we changed the length of the Left-FM and Right-FM electrodes from 50 to 200, keeping the width and height to 5. The quick analysis of the spatial correlation factor indicated that for MTJMSD of 50 atom length, the molecules were strongly correlated with the magnetic moment of the Left-FM and Right-FM (Fig. 7a); however, for 200 atomic length MTJMSDs, molecules were only correlated to the FM electrodes near the junction area (Fig. 7b). Similarly, we also increase the thickness of each FM electrode from 5 to 25, while the length and width were fixed to 50 and 5, respectively. Spatial correlation data for the extreme case of thickness = 25 suggest that Left-FM and Right –FM electrodes were weakly correlated with the molecules' magnetic moment. However, unlike 200 atomic length MTJMSD, the spatial correlation factor was relatively uniform over the whole MTJMSD for 50 atoms thick MTJMSD (Fig. 7c).

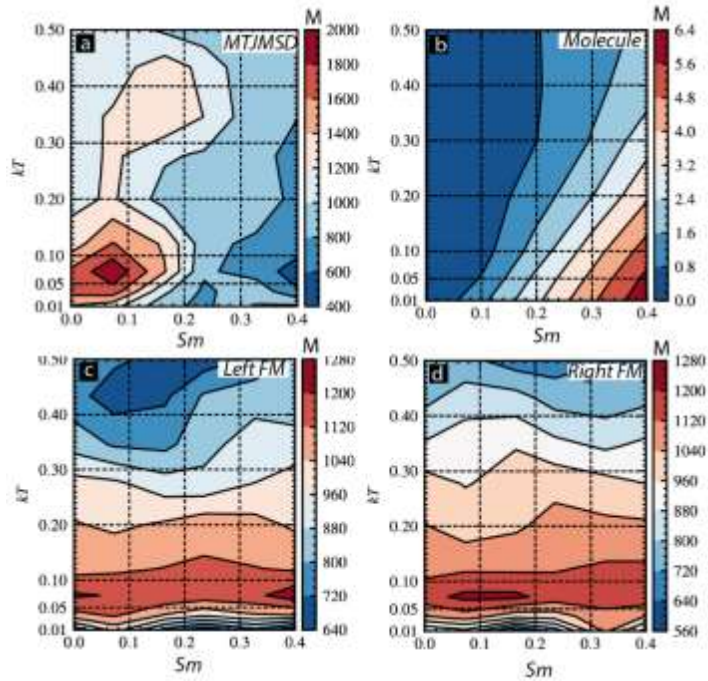

**FIG. 6** Contour plot showing magnetic moment for thermal energy ($kT$) and $S_m$ for (a) full MTJMSD, (b) molecular layer, (c) Left-FM, and (d) Right-FM. For all the cases molecular coupling was $J_{mL} = -1$ and $J_{mR} = 1$.

For further investigation, we plotted the magnetic moment of the MTJMSD and two FM electrodes as a function of the electrode length (Fig. 7d).  The effect of molecule-induced strong exchange coupling could force the large area of MTJMSD only for short lengths (Fig. 7d). As length doubled, MTMSD's Left-FM and Right FM electrode stop aligning perfectly antiparallel to each other, and many metastable phases started becoming possible. As length increased to 150, the MTJMSD magnetic moment was in between the Left-FM and Right-FM electrodes (Fig. 2c). It is apparent that as the length of the electrode increases to 150 or more, FM electrodes appear to have multiple phases leading to lowered magnetic moment (Fig. 7d). Since increasing length did not allow the antiparallel alignment of the two FM electrodes over the full length hence MTJMSD's net magnetic moment was significantly high. The increase in thickness of the FM electrode was more influential in determining the $S_m$ effect on MTJMSD (Fig. 7e). Generally, increasing thickness forced MTJMSD to settle in a higher magnetization state above the individual FM electrode's magnetic moment (Fig. 7e). Interestingly, for the 20-atom thick FM electrode thickness, the MTJMSD's magnetic moment was consistently below the FM electrode magnetic moment. Each data point in Fig. 7d-e was repeated five times, and simulations were conducted for 2 billion iterations to ensure we reach an equilibrium state. We hypothesized that changing the dimensions of the FM electrode impacted the stabilization dynamics; for the 20-atom thick FM electrode thickness, the equilibrium magnetization state was akin to 5 atoms thick FM electrode (Fig. 7e). The size effect data shown in Figure 7 provide direct insights into the consequences of varying the FM electrode dimensions.

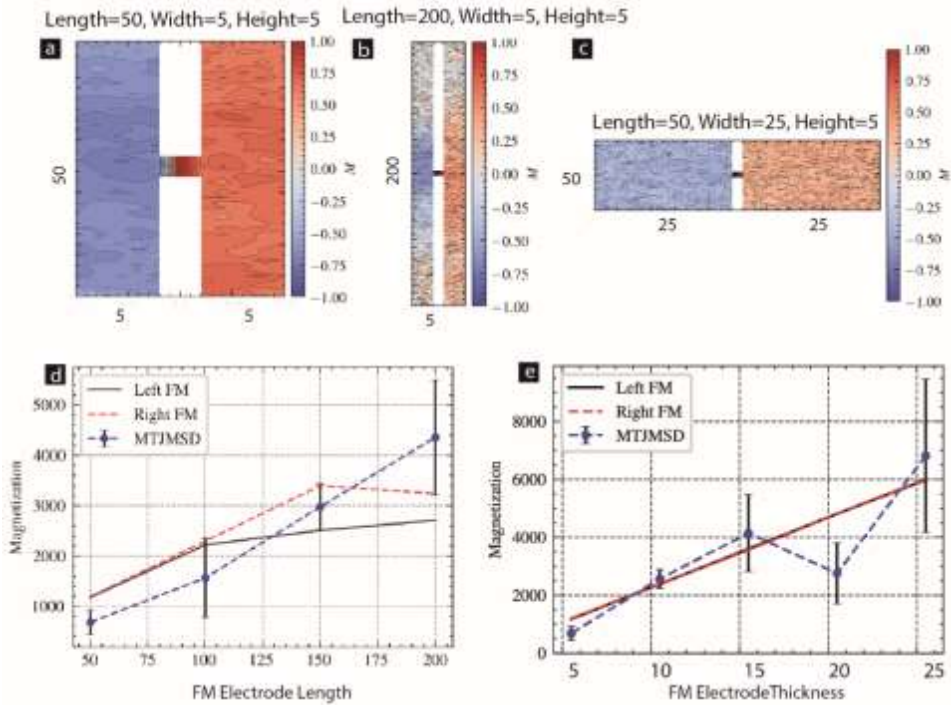

**FIG .7** Spatial correlation factor for FM electrodes with (a) Length = 50, Width=5, and Height =5 (b) Length = 200, Width=5, and Height =5, (c) Length = 50, Width=25, and Height =5. (d) Magnetization vs FM Electrode length (e) Magnetization vs. FM electrode thickness.  For all the cases $S_m$ = 0.2. $kT$ = 0.1, $J_{mL}$ = -1 and $J_{mR}$ = 1. Here, width =thickness of FM electrodes.

## IV. CONCLUSION

We conducted a Monte Carlo simulation to study the impact of the molecular spin state ($S_m$) on the MTJMSD and ferromagnetic electrodes. This research produced a number of lessons that may help in understanding and designing futuristic molecular spintronics devices. (1) In the strong coupling regime, the molecular spin state must be above 0.2 to create a random state to antiparallel FM electrodes in an MTJMSD. (2) Switchable MTJMSD is only possible for a low molecular spin state. (3) In a strong ferromagnetic coupling regime, increasing molecular spin state to S=4 enabled fast equilibration and enhanced the thermal stability molecule induced high magnetic moment. (4) Magnetic electrode thickness and length are critical in determining the molecular spin state effect. We will focus on studying the effect of spin fluctuations on MTJMSD with different molecular spin states in future work.

### ACKNOWLEDGEMENT

This research is supported by National Science Foundation-CREST Award (Contract # HRD- 1914751), Department of Energy/ National Nuclear Security Agency (DE-FOA-0003945). Author contributions: Andrew Grizzle conducted simulations studies. Andrew Grizzle developed analysis software to analyze the data and Christopher D'Angelo wrote C++ program under supervision of Pawan Tyagi. and Andrew Grizzle wrote the manuscript and analyzed the data.

### DATA AVAILABILITY STATEMENT

The data that support the findings of this study are available from the corresponding author upon reasonable request.

# Supplementary Material

## Spin State of Single Molecular Magnet (SMM) Creating Long Range Ordering on Ferromagnetic Layers of Magnetic Tunnel Junction -A Monte Carlo Study

Andrew Grizzle, Christopher D'Angelo, Pawan Tyagi*
Center for Nanotechnology Research and Education, Mechanical Engineering, University of the District of Columbia, Washington DC-20008, USA
Email of corresponding author: ptyagi@udc.edu

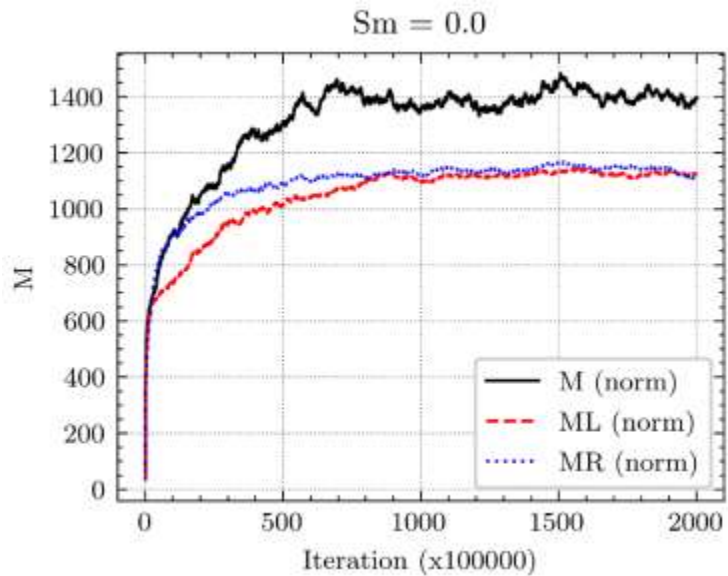

Fig. S1: M vs. time for Sm=0.

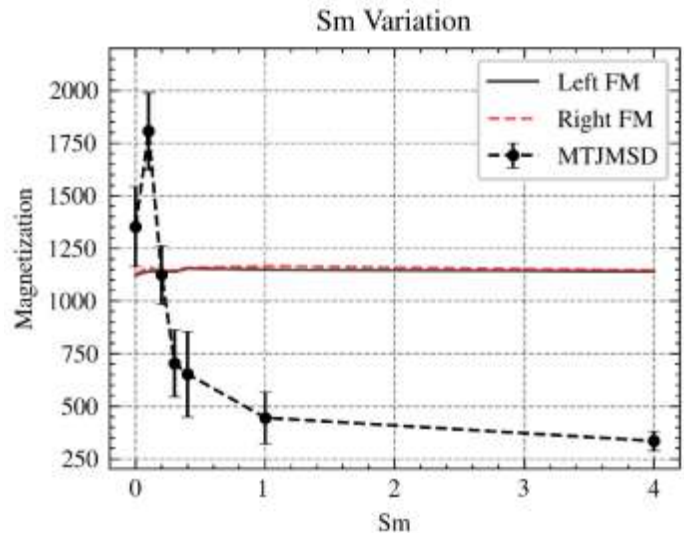

Fig.S2: Magnetization vs. Sm for 0-4 Sm range

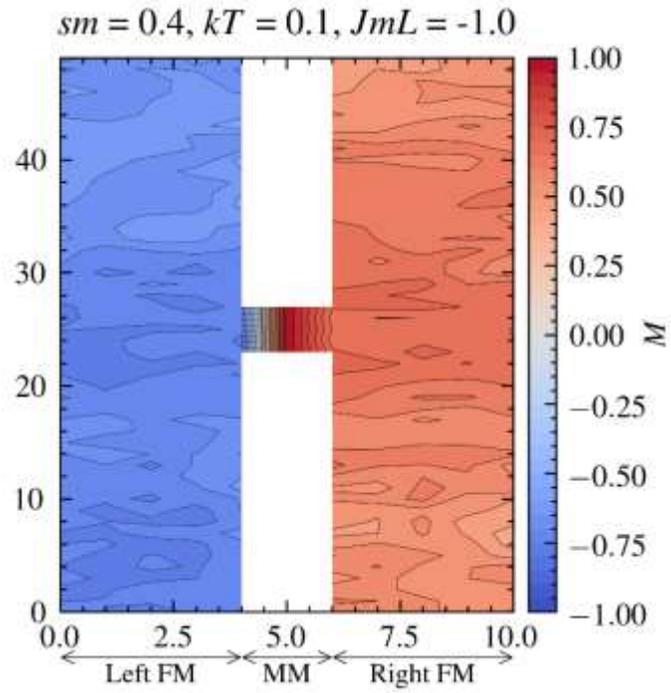

Fig. S3: Spatial correlation factor for Sm=0.4

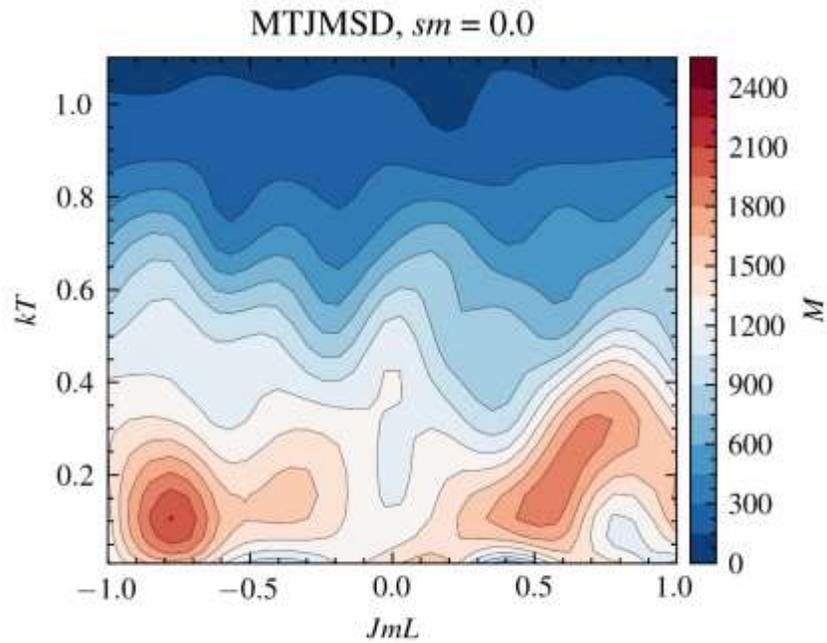

Fig. S4: kT vs JmL for Sm=0. JmR=|JmL|.